\documentclass [floatfix, showpacs, aps, pra, twocolumn, 10pt]{revtex4-1}

\usepackage{amsmath}
\usepackage{graphicx}
\usepackage{bbm}
\usepackage{hyperref}
\usepackage{amssymb}
\usepackage[english]{babel}
\usepackage[latin1]{inputenc}
\usepackage{lmodern}
\usepackage[T1]{fontenc}
\usepackage{amssymb}
\graphicspath{{Figs/}}
\begin{document}

\title{Simultaneous Detection of H and D NMR Signals in a micro-Tesla Field}

\author{Giuseppe Bevilacqua}
\affiliation {DIISM, University of Siena - Italy}

\author{Valerio Biancalana}
\affiliation{DIISM, University of Siena - Italy}

\author{Yordanka Dancheva}
\affiliation{DSFTA, University of Siena - Italy}

\author{Antonio Vigilante}
\affiliation{DSFTA, University of Siena - Italy}

\author{Alessandro Donati}
\affiliation{DBCF, University of Siena - Italy}

\author{Claudio Rossi}
\affiliation{DBCF, University of Siena - Italy}

\begin{abstract}

We present  NMR  spectra of  remote-magnetized  deuterated water, detected in an unshielded environment  by means  of  a differential  atomic magnetometer.  The  measurements are performed in a $\mu$T field, while pulsed techniques are applied --following the sample displacement-- in a 100~$\mu$T field, to tip both D and H nuclei by controllable amounts. The broadband nature of the detection system enables  simultaneous detection of the two signals and accurate evaluation of their decay times. The outcomes of the experiment demonstrate the potential of ultra-low-field NMR spectroscopy in important applications where the correlation between proton and deuteron spin-spin relaxation rates as a function of external parameters contains significant information.

\vskip 1cm  
  
\includegraphics [width=12cm] {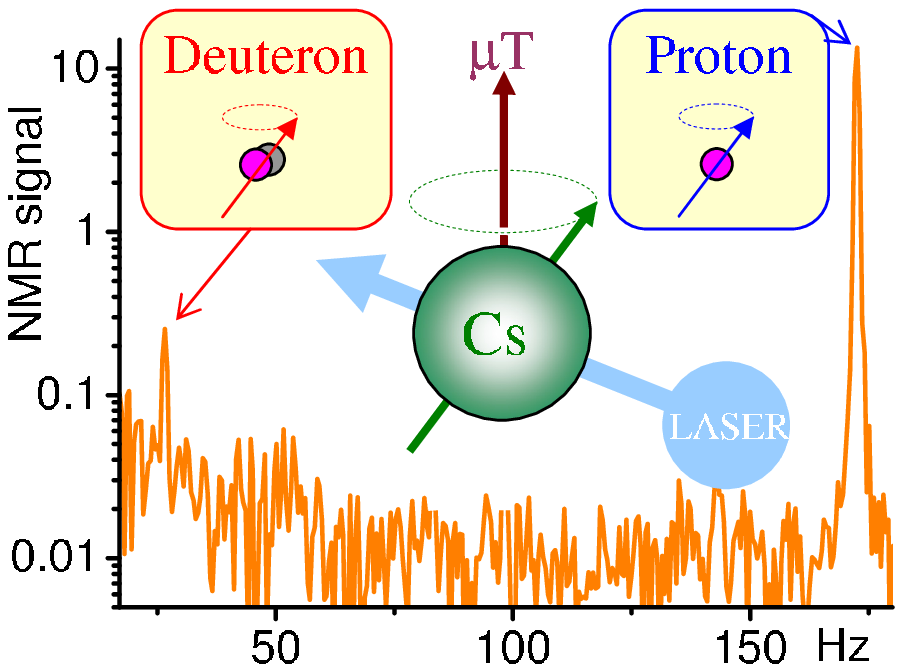}
\centering

\end{abstract}

\maketitle

\section{Introduction}

Experimental NMR relaxation parameters such as spin-lattice relaxation rates, $R_1$, spin-spin relaxation, $R_2$, and spin-lattice relaxation in the rotating frame, $R_{1\rho}$, show a dependence on the fluctuation of local magnetic fields coupled with the  nuclear magnetic moment observed. While proton relaxation in diamagnetic systems is dominated by fluctuations in the dipole-dipole interactions between proton nuclei and its magnetic environment, deuteron relaxation mechanism is dominated by fluctuations in the electric field gradient around the nucleus. Both $R_1$ and $R_2$ of proton and deuteron were used to study the dynamic and structural properties of complex biosystems in solution \cite{palmer_cr_04}. $R_1$ relaxation rates for proton and deuteron are both dependent on the spectral densities $J(\omega)$ of the fluctuating dipolar and quadrupolar interaction at $\omega=\omega_L$ and $\omega=2\omega_L$ respectively, $\omega_L$ being the Larmor angular frequency. The spin-spin relaxation, $R_2$, is also affected by the low-frequency components of the spectral density. This behaviour makes $R_1$ suitable for the investigation of fast and intermediate dynamic processes, while $R_2$ is more sensitive and better suited to studying slow motion processes (where, in the case of high-field conventional NMR, {\it fast} means $\tau_c < 10^{-11}$s, {\it intermediate} $\tau_c $ is  in the range of  $10^{-10} \div 10^{-9} $s, and {\it slow} motion is when $\tau_c > 10^{-9}$s). In the past, the deuteron transverse relaxation rate constant, $R_2$, was used to analyse the slow motion behaviour of phospholipid bilayer membranes \cite{bloom_bc_87} and dendrimer macromolecules \cite{metzer_mrm_92},  as well as to discriminate between tissues under physiological and pathological conditions \cite{block_mrm_87, ababneh_mrm_05}. Similarly, proton spin-spin relaxation has been investigated in the study of slow frequency dynamics in complex systems \cite{case_acr_02, barbieri_pol_98, akke_jac_96}.

Both proton and deuterium spin-spin relaxation rates have been investigated to quantify protein dynamics \cite{kay_bcb_98}, pore and collagen-bound water in cortical bone and cartilage \cite{ong_jbmr_12}, and water behaviour in mesoporous materials \cite {wehrli_jmr_13}, as well as for other scientific and technological applications \cite{reiter_mrm_09, steiner_jpca_11, huang_mrm_13, birdwell_ef_15, raj_plosone_14, cabrales_amr_08, meng_bt_13}. 
A relationship exists between proton and deuteron relaxation rates and is best investigated if both sets of measurements are obtained using the same sample and under the same experimental conditions. 
 In other words, any possible correlation between proton and deuteron spin-spin relaxation rates must result from experiments in which the measurements of  $R_2$  of both proton and deuteron can be determined simultaneously.

 \begin{figure}[htbp]
   \centering
  \includegraphics [width=8cm] {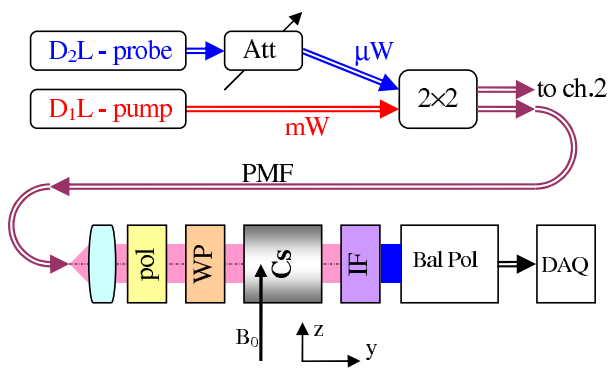}
  \caption{One channel of  the magnetometer. Pump (D$_1$L) and (attenuated) probe (D$_2$L) radiations are mixed ($2 \times 2$) and fiber coupled to the Cs cells. Suited optics and waveplates (polarizer (pol) and waveplate (WP)) provide a collimated beam with opportunely polarized radiations. Following interaction with the Cs vapour, the pump light is stopped by an  interference  filter (IF), while the Faraday rotation of the probe polarization is measured by a balanced polarimeter (BalPol) and digitized (DAQ). The magnetometer operates in a highly homegeneous transverse field ($B_0$) in the $\mu$T range.}
  \label{oursetup}
\end{figure}

 In the present paper we  show a new method for the  determination of $R_2$ of both proton and deuteron nuclei, using an apparatus for nuclear magnetic resonance at ultra-low magnetic fields. 
 We take advantage of the potentialities of NMR detection at ultra-low fields to register signals
 containing both D and H contributions and make use of non-inductive sensors with a flat response over the whole frequency range of interest.

Ultra-low field NMR has been performed using superconducting quantum interference devices (SQUIDs) for almost three decades \cite{pines_prl_83} and with atomic optical magnetometers (AOMs) for more than one \cite{romalis_prl_05}. Compared to SQUIDs, AOMs have the advantage of not requiring cryogenics, which facilitates the sensor-sample coupling and simplifies the setups. Our efforts have been concentrated on developing robust setups operating in unshielded environments. This choice further simplifies the apparatus, at the expense of sensitivity levels about one order of magnitude worse than achievable with similar AOMs having the same bandwidth but operating in shielded volumes.

This work has been accomplished with an apparatus in which samples are remotely magnetized and pneumatically shuttled to the sensor head. Selective tipping of individual spin species is accomplished reinforcing the bias field $B_0$ for the duration of the tipping pulse. Once rotated by certain amount the spins precess at frequencies as low as few tens of Hz and are detected using a broadband non inductive sensor.

\section{Magnetometer experimental set-up}

The non-inductive detection is based on an atomic magnetometer using optically pumped Cs vapour \cite{bevilacqua_apb_16}. The AOM used has two identical arms enabling the application of a differential technique to reject the ambient noise. Each  arm (see Fig.~\ref{oursetup}) contains a gas buffered Cs cell illuminated by circularly polarized laser light resonant with the $D_1$ line (pump laser) and by weak linearly polarized laser light nearly resonant with the $D_2$  line (probe laser). The  two  radiations  are collimated  into a beam --10~mm in diameter-- propagating along the $y$ direction, while a homogeneous magnetic field in the $\mu$T range is transversely oriented along $z$.

The AOM consists of a Bell\&Bloom \cite{bellandbloom61} apparatus where the optically induced magnetization of the atomic vapour precesses around a magnetic field made of a dc bias superimposed on a much weaker field under measurement which is produced by the precessing nuclei (more details are given in Refs.\cite{bevilacqua_jmr_16, bevilacqua_apb_16}). The bias field is accurately controlled in terms of intensity, direction and inhomogeneities. 

The polarimetric signal consists of a nearly sinusoidal term, whose frequency $f_0$ is the modulation frequency of the pump laser wavelength, and is resonant with the atomic precession frequency. Any change in the bias field results in a slight phase variation of the polarimetric signal, similarly a damped harmonic oscillator driven by a near resonant signal. 

The total phase $\theta(t)=2\pi f_0 t + \varphi(t)$ of the polarimetric signal (which  corresponds  to  the atomic Larmor precession angle)  is inferred and the phase $\varphi$, which is weakly dependent upon time ($\dot \varphi \ll f_0$), makes it possible to determine the Faraday rotation associated with the time-dependent magnetic field within a bandwidth spanning from dc up to a few hundred Hz. 

\begin{figure}[htpp]
\includegraphics [width=8cm] {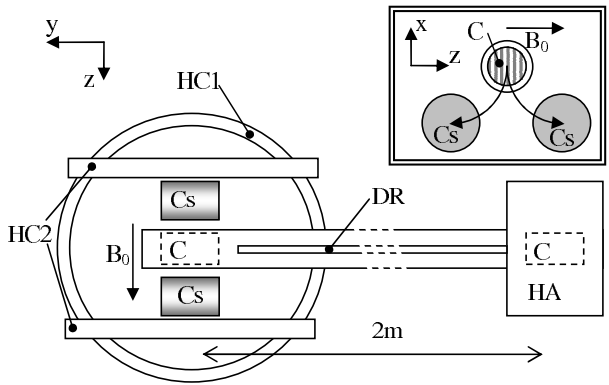}
\centering
\caption{Sample premagnetization, shuttle, positioning, and detection geometry. The sample cartridge C is premagnetized in a Halbach array (HA) and then pneumatically shuttled to the proximity of the sensors (Cs), 2~m apart. The sample magnetization adiabatically follows the local field direction as long as the precession is fast. To this end a square coil (DR) in the xy plane provides a driving field during the shuttling. At the arrival, two secondary Helmholtz pairs (HC1, HC2) provide extra fields to manipulate the nuclear spin and (if necessary) to reinforce the bias field, respectively. The inset shows a front view of the sample and sensors, with the geometry maximizing the differential signal of the dipolar sample field.}
\label{fig:nmrsetup}
\end{figure}

Common-mode field variations generated by distant noise sources induce equal changes in the phases $\varphi_1, \varphi_2$ measured at the two arms, while local sources (C in the inset of Fig. \ref{fig:nmrsetup}) produce difference-mode field variations, which appear with opposite signs. Thus a differential (gradiometric) measurement provides a common-mode free signal $\Delta \varphi = \varphi_1 - \varphi_2$ driven by the nuclear precession in the sample.

\begin{figure}[htpp]
\includegraphics [width=8cm] {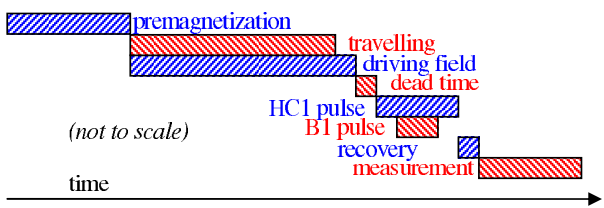}
\centering
\caption{Timing  of  the  cycle  for NMR  spectroscopy. The sample is premagnetized for about 5 s. Then the driving field is turned on and the sample is shuttled. After its arrival the driving field goes off. After a few ms dead time, the bias field $B_0$ is reinforced by a rectangular pulse on HC2, and a resonant tipping pulse is applied (via CH1). About 30~ms are then necessary to allow the magnetometer to reach steady operation then data acquisition may start. Finally, the cartridge is pulled back to the Halbach array, making the system ready to start the next cycle.}
\label{fig:timing}
\end{figure}

The two arms of the magnetometer are displaced from each other by a baseline of $5.6~\mbox{cm}$  along the $z$ direction. The magnetometer resonance line has a linewidth of the order of ~30~Hz and ensures a sensitivity of $100~\mbox{fT}/\sqrt{\mbox{Hz}}$ in the frequency range $10 \div 150$~Hz.

\section{NMR spectroscopy experimental set-up}

The experimental  set-up for nuclear spin polarization, transport and manipulation is represented in Fig.\ref{fig:nmrsetup}. The  sample to  be  analyzed is contained in  a  5~cm$^3$ cartridge. It is first magnetized in a pre-polarization device based on  Nd permanent magnets arranged in a Halbach array providing a 1~T magnetic field in a volume
25~mm in diameter and 50~mm in length. Subsequently, a driving field is turned on and the sample is pneumatically shuttled \cite{biancalana_rsi_14} to the AOM  head.  The travel time is about 140~ms, after which the driving field is switched off, and the tipping pulses are applied. Tipping can be achieved either non-resonantly (non-selectively) by applying a rectangular pulse as a transverse field or resonantly. The measurements shown in this paper are obtained with the resonant technique. 

The measurements  were performed using an automated cycling procedure, whose timing is sketched in Fig.~\ref{fig:timing}. The magnetic field inhomogeneities along the sample displacement path are small enough so that the proton magnetization follows the field adiabatically. A driving field is necessary to ensure that polarization of the D nuclei is also preserved. This field is switched off adiabatically after the transfer of the sample, so that the magnetization is eventually aligned with the static field.

A webcam detects the actual position reached by the sample with  respect  to  the  sensors shot-by-shot, making it possible to automatically disregarding bad shots (of which a small percentage nonetheless remain).

 \begin{figure}[htpp]
\includegraphics [width=8cm] {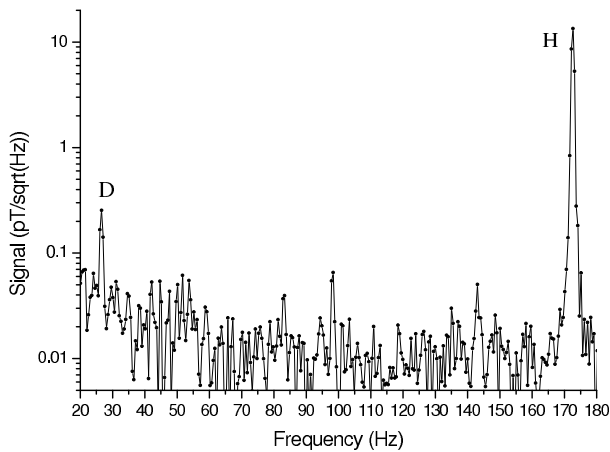}
\centering
\caption{A simultaneous spectrum of H and D nuclei. The data used to evaluate this spectrum are obtained as an average of 1000 shots. A selective (cosine enveloped) dual frequency tipping pulse is applied to rotate both H and D nuclei by $\pi/2$. The data are recorded in a dc field of $4~\mu~T$, producing 172~Hz precession frequency for the proton and 26~Hz for the D nuclei. The tipping pulse is applied simultaneously to a HC2 pulse, enhancing the dc field up to 140~$\mu$T, so that in spite of its short (20~ms) duration, it  contains 120 periods for H and 20 periods for D. Technical noises (e.g. from the mains at 50~Hz) are identified and subtracted. High magnetic field homogeneity in the detection region provides very low instrumentation bandwidth. H decay time of 1.2~s is measured.}
\label{fig:spectrum}
\end{figure}

\section{Results}

The spectrum plotted  in Fig.~\ref{fig:spectrum} is obtained with a sample containing 2~ml D$_2$O and 2 ml H$_2$O, by averaging over 1000 traces. The difference in amplitude between the two peaks is due to the shorter decay time and the smaller gyromagnetic factor of D. The latter causes a lower initial magnetization level compared to H, although this has no relevance at the detection stage, thanks to the non-inductive nature of the sensor.

\begin{figure}[htpp]
\includegraphics [width=8cm] {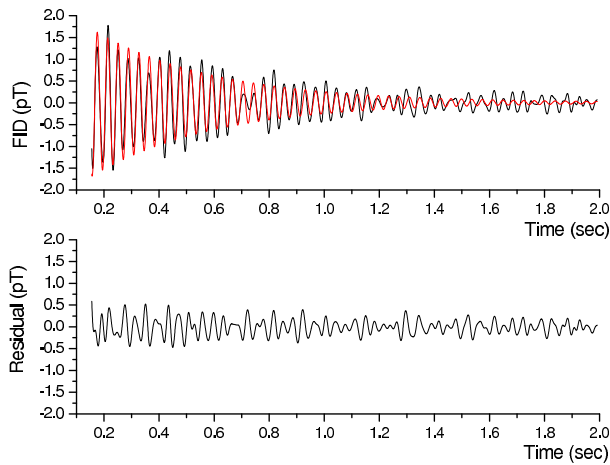}
\centering
\caption{Deuteron signal as a function of time as extracted from the time trace by means of a linear band-pass filter ($4^{th}$ order, 20 Hz bandwidth). A decay time of 0.4~s is estimated. The lower plot shows the residual from a best fit targeted to a damped oscillation.}
\label{fig:Dfid}
\end{figure}

The decay times of both H and D nuclei are estimated from the averaged traces on the basis of discrete Fourier transform methods \cite{duda_intech_12}, as well as using non-linear fitting procedures targeted to damped oscillation and in both cases values of 1.2~s and 0.4~s are obtained, respectively. The upper plot in Fig.\ref{fig:Dfid} shows the time-domain D trace, as obtained with the application of a band-pass filter, and the corresponding fitting curve. Their difference is shown in the lower plot.

In conclusion, we have demonstrated that a ultra-low field NMR experiment based on a differential optical magnetometer enables the detection of proton and deuteron precession signals in a micro-Tesla field in an unshielded environment. The possibility of detecting H and D signals simultaneously at various intensities of the magnetic field in the ultra-low field range makes this research of interest in the field of biochemistry and in other applications where relaxation mechanisms with correlation frequencies in the sub-kHz range are expected. Such long correlation time measurements render this ultra-low-field magnetometry a complementary method to the conventional high-field NMR.

\acknowledgments{The authors thank E.Thorley and "Lingo Lingo Sprachdienst" for revising the English of the manuscript.}

\appendix

\section{Supplemental material}
The following appendices contain supplemental material completing the information provided in the letter  "Simultaneous Detection of Hydrogen and Deuterium NMR Signal in a micro-Tesla Field". Additional details  about the magnetometer sensor (atomic medium preparation and Faraday rotation detection), sample shuttling system, sample premagnetization assembly, data acquisition and analysis are briefly presented.


%
%

\section{Atomic medium and laser radiations}
Atomic magnetometry dates several decades \cite{bell_prl_61} or, in some sense, even more than one century \cite{macaluso_98}, but it is having a revival in the last two decades, thanks to advances in solid state laser technology. A review in this field can be found in Ref.[\cite{budker_nat_07}]. 

\begin{figure}[htbp]
   \centering
  \includegraphics [width=\columnwidth] {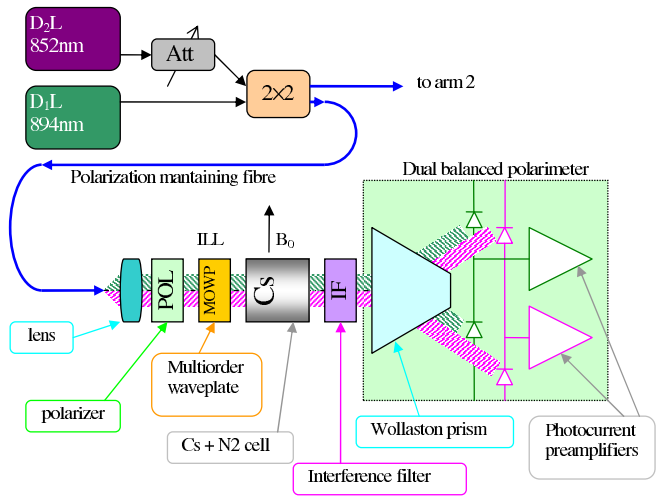}
  \caption{Schematics of  a single arm of  the magnetometer. The
    D$_1$L  radiation pumps the  atoms and  the D$_2$L one probes them.  The two  radiations are injected in polarization maintaining optical fibres and merged by 2 input / 2 output 
    ($2\times2$)  mixer. Then, they are coupled to the  sensors  by other  polarisation-maintaining fibres. At the output, an illuminator collimates the two radiations into a single beam and makes their polarisations appropriate (renders circular the polarisation of the pump, while leaving linear the  probe one). The Cs cell  is thus illuminated  by  the two
    co-propagating  radiations.    After the interaction with the Cs vapour, the pump radiation is stopped
    by  an  interference  filter (IF), while the probe is analysed by means of a Wollaston polariser oriented at $45^\circ$ with respect to the input polarisation plane, in the so-called balanced  polarimeter configuration. Such polarimeter measures the orientation of the probe polarisation plane for the two halves of the beam separately. The magnetometer works as a gradiometer of different bases when acquiring the Larmor signals coming from the two cells (long baseline) or the two halves of a beam of the same cell (short baseline).}
  \label{fig:1arm}
\end{figure}

The magnetometer used in this experiment is an updated version of a setup previously described \cite{biancalana_ap_16, biancalana_arnmrs_13, belfi_josa_09}. It consists of two identical arms having as sensors Cs vapour and $N_2$ buffer gas (at 23 Torr) in sealed cells.
Each  arm (see Fig.~\ref{fig:1arm}) contains an illuminator  providing  two co-propagating laser radiations, a Cs sensor, and a balanced polarimeter. The polarimeter analyzes the polarization of the emerging probe radiation in half-beam sections so that, as a total, four polarimetric signals are available. 

The two laser radiations have different wavelengths. 
One of them is resonant with the D$_1$ transitions of  the Cs atoms at 894~nm. This light is kept at a relatively high intensity (mW/cm$^2$) and it is used to optically pump the
atoms. The other radiation is near resonant with the D$_2$  transitions at 852~nm and it is used to probe the atomic spins precession. 

The magnetometric head is fibre-coupled to the laser sources. More precisely, the pump radiation  D$_1$L and the probe radiation D$_2$L are
coupled to polarisation-maintaining fibres.  A  proper mixture of the two radiations is then accomplished using an on-fibre mixer. Both the light sources are based on solid-state continuous-wave devices. 

The probe  radiation  is tuned at a constant frequency (the optimum is found to be at about 2~GHz on the blue wing of the  $^2S_{1/2}\rightarrow ^2P_{3/2}$ (D$_2$) unresolved resonance set), and is attenuated before mixing by means of an on-fibre device. The pump  radiation is used at full power (in the mW range), and is broadly frequency modulated, in such way that it goes periodically in- and out-of resonance with the $F_g=3\rightarrow F_e=4$ component of the $^2S_{1/2}\rightarrow ^2P_{1/2}$ (D$_1$) transition set. 

At the fibre exit, just before the vapour cells to be illuminated, the  two  laser lights are collimated  into  parallel beams   about  10~mm in  diameter, their linear polarisation are reinforced by a polarising cube, and finally a special  multi-order waveplate (MOWP) makes  the pump radiation
circularly polarised, while leaving the probe one linearly polarised. This technique is based on an advantageous opportunity offered by the two-colour operation of the system. In fact, it is possible to build a multi-order quartz wave-plate with a nominal relative delay of 4.75 wavelengths for  D$_1$ radiation and 5.00045 wavelengths for  D$_2$ radiation: such plates when oriented at $45^\circ$ with respect of the polarization direction of linearly polarized radiations renders circular only the component that is subject to a $\lambda/4$ dephasing.

\section{Faraday Rotation Detection}
The pump laser aligns the Cs atomic spins along its propagation axis producing a macroscopic magnetization of the atomic vapour. The presence of a perpendicular magnetic field forces the atomic magnetization to precess in the orthogonal plane, so that its component along the optical axis oscillates in time with the Cs Larmor frequency.
In order to counteract the relaxation processes, the pump laser is made periodically resonant, synchronously with the Larmor precession.
Upon these conditions, the atomic sample behaves similarly to a damped oscillator resonantly forced. In particular the Cs magnetization along the optical axis evolves with a phase, which depends on the detuning between the the Larmor frequency and the  the pump laser modulation frequency (forcing signal).

The measure of such phase provides  direct and accurate information about the actual field. The latter results from the superposition of a dominant bias term and a small time-dependent component generated by the nuclear spins under measurement.

The precession of the atomic magnetization can be detected with a high efficiency looking at the circular birefringence of the medium, which is particularly strong for a radiation near resonant with the atomic system, as for the case of the probe radiation. Thus, the polarimetry on the transmitted probe radiation offers an efficient tool for deriving the time dependent magnetic field acting on the Cs atoms, i.e. the time evolution of the nuclear spins in the NMR sample.

\section{Sample shuttling system, and premagnetization}
A pneumatic sample shuttling system is set up to displace the sample from the premagnetization area to the measurement area. It uses compressed air, at about 0.4~MPa, controlled by standard electrovalves which are computer controlled with a 1~ms time accuracy. Typical shuttling times are 140~ms for forward displacement and 1.2~sec for backward displacement, for a pipe of 2~m in length. Shorter times (down to about 50~ms) can be achieved using higher pressure, at expense of some reduction in reliability and in shot-by-shot reproducibility. At the end of the fast, forward displacement, the sample is pneumatically decelerated, in order to avoid bouncing and mechanical shock at the arrival. 

The premagnetization of the sample is accomplished in a home-made Halbach assembly using sixteen 1-inch, cubic Nd magnets. The magnetic field strengths inside the magnetizing volume is  approximately  1~T. Such volume has a cylindrical shape,  25~mm in diameter and 50~mm in length. The diameter matches the outer diameter of the shuttle pipe, and the length exceeds the sample-holder cartridge length. The cartridge is made of polyether-ether-ketone polymer (PEEK), a material selected for its excellent mechanical and chemical resistance properties. The cartridge is accurately machined to match the inner pipe diameter and may host up to a 5~ml sample.  Details about the shuttling system and the premagnetization assembly are available in a technical note \cite{biancalana_rsi_14}.

\section{Data analysis}
The polarimetric signals from the four channels of the magnetometer are acquired using a standard ADC National Instruments card (NI6250) with 1~MS/s rate per channel and $2^{20}$ samples. The phase is extracted on the basis of  numerical  Larmor frequency demodulation, and it is converted in magnetic field units according to the expression given in the Appendix of Ref.[\cite{biancalana_ap_16}]. The field-difference between two channels of the magnetometer (the baseline is about 55~mm) is analyzed. The field-difference signal contains signal due to spin nucleus free decay (FID signal), while disturbances originating from far located sources appear as common-mode terms and are cancelled by subctraction. The NMR spectrum is evaluated from the fast Fourier transform of the differential field signal. In order to visualize FID signal coming from a specific nuclear spin a linear numerical filtering (band-pass filter of a specific order) is applied.


\end{document}